\documentclass[preprint] {article}
\usepackage{amsmath}
\usepackage{amsfonts}
\usepackage{amssymb}
\usepackage{geometry}
\usepackage{graphicx}
\usepackage{braket}
\allowdisplaybreaks
\newcommand {\bp}{\begin{pmatrix}}
\newcommand {\ep}{\end{pmatrix}}
\newcommand{\be}{\begin{equation}} \newcommand{\ee}{\end{equation}}
\newcommand{\bea}{\begin{eqnarray}}\newcommand{\eea}{\end{eqnarray}}

\begin{document}
\title{Taming Hamiltonian systems with balanced loss and gain via
Lorentz interaction : General results and a case study with Landau
Hamiltonian}

\author{ 
Pijush K. Ghosh\footnote {{\bf email:} 
pijushkanti.ghosh@visva-bharati.ac.in}} 
\date{Department of Physics, Siksha-Bhavana, \\ 
Visva-Bharati University, \\
Santiniketan, PIN 731 235, India.}
\maketitle

\begin{abstract}
The kinetic energy term of Hamiltonian systems with balanced
loss and gain is not semi-positive-definite, leading to instabilities
at the classical as well quantum level. It is shown that an additional
Lorentz interaction in the Hamiltonian allows the kinetic energy
term to be semi-positive-definite and thereby, improving the stability
properties of the system. Further, a consistent quantum theory admitting
bound states may be obtained on the real line instead of Stoke wedges on
the complex plane. The Landau Hamiltonian in presence of balanced loss and
gain is considered for elucidating the general result. The kinetic energy
term is semi-positive-definite provided the magnitude of the applied external
magnetic field is greater than the magnitude of the `analogous magnetic field'
due to the loss-gain terms. It is shown that the classical particle
moves on an elliptical orbit with a cyclotron frequency that is less than its
value in absence of the loss-gain terms. The quantum system
share the properties of the standard Landau Hamiltonian, but, with
the modified cyclotron frequency. It is shown that the Hall current has
non-vanishing components along the direction of the external uniform
electric field and to its transverse direction. The Pauli equation in
presence of balanced loss and gain is shown to be supersymmetric.
\end{abstract}

\tableofcontents
\vspace{0.3in}

\section{Introduction}

Hamiltonian systems with balanced loss and gain have received
considerable interest\cite{bpeng,ben,ben1,ivb,sagar,ds-pkg,khare,
pkg-ds,ds-pkg1, p6-deb} in the literature due to its potential
applications in various branches of physics. One of the major
drawbacks of such Hamiltonian systems is that the kinetic energy
term is not positive-definite. Consequently, stable classical
solutions are obtained for specific choices of the potentials and
that too, within restricted ranges of parameters. Further, a consistent
quantum theory for such systems admitting bound states requires extensions
of the eigen-value problem to the complex domain. The normalizability
of eigenfunctions is obtained in specific Stoke wedges. Although such
an extension of quantum mechanics is consistent from the viewpoint of
axiomatic foundation of the subject\cite{ben-review,ali-review}, no
experimental support for such theories has been observed so far. 
A natural question one would like to pose at this juncture is whether or
not a modification of Hamiltonian systems with balanced loss and gain is
possible so that the kinetic energy term can be made semi-positive definite.

The purpose of this article is to show that the answer to the question posed
above is in the affirmative and can be achieved by the inclusion of  Lorentz
interaction in the system. It may be noted in this regard that
a vanishing trace of the matrix appearing in the quadratic form of momenta
defining the kinetic energy term necessarily makes it non-positive-definite.
All previous investigations on Hamiltonian systems with balanced loss and gain
belong to this class\cite{pkg-ds,ds-pkg1,p6-deb}. It is shown that a
non-vanishing trace of this matrix
requires the inclusion of non-dissipative velocity-dependent forces in the
system with asymmetric coupling between the gain and loss degrees of freedom.
The Lorentz force belongs to this class and its inclusion in the system
raises the possibility of making the kinetic energy term positive-definite for
some ranges in the parameter space. It is worth emphasizing here that the
Lorentz force appears in diverse areas of science and the working principle
of many devices are based on it. Thus, inclusion of Lorentz force in the study
of Hamiltonian system with balanced loss and gain is only natural and may
open up new avenues for theoretical understanding of these systems with
possible technological applications.

The generic formulation of Hamiltonian system with space-dependent balanced
loss and gain is modified appropriately to include Lorentz force in the system.
This has been done for arbitrary number of particles and also for generic
space-dependence of the gain-loss co-efficients. It has been shown that
the inclusion of Lorentz force implies that the balancing of loss and gain
is not necessarily in a pair-wise fashion and it may be achieved in several
ways. This should be contrasted with all previous investigations on Hamiltonian
systems with balanced loss and gain, where balancing is necessarily achieved
in a pair-wise fashion. This allows more flexibility in constructing and
controlling Hamiltonian systems with balanced loss and gain.

It is shown on general ground that Hamiltonian systems with balanced loss-gain
and in presence of Lorentz force may be interpreted as defined in the
background of a metric subjected to an effective external magnetic field,
whose magnitude depends on the applied magnetic field related to the Lorentz
force and an analogous magnetic field due to the loss-gain terms. The
background metric is either Euclidean or pseudo-Euclidean, depending on the
region in the parameter space in which it is defined. The metric is Euclidean,
if the magnitude of the applied magnetic field is greater than the magnitude
of the analogous magnetic field and pseudo-euclidean, otherwise. The specific
signature of the pseudo-euclidean metric is model dependent. For the case of
zero applied magnetic field, the known result that the background metric is
only pseudo-Euclidean is recovered.

The Landau Hamiltonian\cite{landau} appears in the description of various
physical phenomena, including quantum Hall effect and spintronics based devices.
The Landau Hamiltonian with balanced loss and gain is considered as an
example to elucidate the general result. There are three regions in
the parameter-space corresponding to an Euclidean, a pseudo-Euclidean and
a negative-definite metric. It is shown that the system for the case of
Euclidean metric admits periodic solution with the cyclotron frequency
determined in terms of the effective magnetic-field, thereby, with a reduced
value compared to the case of Landau Hamiltonian without the loss-gain terms.
Further, the circular orbit of the particle is deformed to an ellipse due
to the presence of the loss-gain terms. The Hamiltonian is bounded from below.
There are no stable solutions for the case of pseudo-Euclidean metric. The
solutions for the case of a negative-definite metric share the same feature
as in the case of Euclidean metric. However, the Hamiltonian is not bounded
from below for this case.

The canonical quantization scheme is followed to obtain quantum Landau
Hamiltonian with balanced loss and gain terms. The parameter space is again
divided into three regions as in the case of corresponding classical system.
For the case of Euclidean metric, all known results of the standard Landau
Hamiltonian are reproduced with a modified expression for the cyclotron
frequency in terms of the effective magnetic field. For the case of
pseudo-Euclidean metric, no bound states are possible either on the real lines
or by extending the problem to complex domain. For the case of negative
definite metric, the Hamiltonian is bounded from above. The eigenfunctions
of the Hamiltonian in this region are also eigenfunctions of the Hamiltonian
for the case of Euclidean metric. However, the energy eigenvalues of the
Hamiltonian in these two regions differ by an overall multiplication factor
of $-1$.

The Hall effect with balanced loss and gain is studied by including an
external uniform electric field to the Landau Hamiltonian. It is shown that
the Hall current has non-vanishing components along the direction of the
applied electric field as well as to its transverse direction. The angle
between the direction of the Hall current and that of the applied electric
field depends on the gain-loss parameter. The result is valid at the classical
as well as quantum level.

It is known that the Pauli equation has an underlying
supersymmetry\cite{khare1}. A similar investigation is carried out
for the Landau Hamiltonian with balanced loss and gain by including an
additional Zeeman interaction term. It is shown that the resulting
system has an underlying ${\cal{N}}=2$ supersymmetry provided Zeeman
interaction contains the effective magnetic field instead of the external
magnetic field. An alternative interpretation is that the Zeeman interaction
still involves the external magnetic field, but, the Land$\acute{e}$
$g$-factor is modified in presence of loss-gain terms. The spectra and the
state-space structure of the resulting Hamiltonian is identical with the
standard Pauli Hamiltonian.

The plan of presentation of the results is the following. The formalism
of the problem along with the general results are described in Sec. 2.
The role of Lorentz interaction for having a semi-positive definite
Hamiltonian is discussed in sub-section 2.1.
The Hamiltonian formulation involves representation of some matrices
which are presented in Sec. 2.2. The Sec. 2.3 contains discussions on
an effective description of the system where loss-gain terms are absent
and the system is subjected to an effective external magnetic field.
The Landau Hamiltonian with balanced loss and gain is introduced in Sec. 3
with the results for the classical and the quantum systems in Sec. 3.1 and
Sec. 3.2, respectively. The Hall effect in presence of balanced loss and
gain is described in Sec. 3.3. In Sec. 3.4, Pauli equation with balanced
loss and gain is shown to be supersymmetric. Finally, the results are
summarized in Sec. 4.
The Appendix-A in Sec. 5 contains representation of matrices for which
balancing of loss-gain terms are not necessarily in a pair-wise fashion.

\section{Formalism \& General result}

Bateman's approach to find a Hamiltonian for a dissipative system is to
embed it in an  ambient space with twice the degrees of freedom of the original
system\cite{bateman}. The extra degrees of freedom constitute an auxiliary
system and the Hamiltonian is obtained for the combined original
plus the auxiliary systems. In this approach, neither the original nor
the auxiliary system alone is Hamiltonian. 
A Hamiltonian formulation for a system with
both loss and gain without the introduction of an auxiliary system needs
a separate treatment from that of Bateman. 
The point may be
explained in terms of a system governed by the equations of motion:
\bea
\ddot{x}_i + \sum_{k=1}^N \eta_{ik}(x_1, \dots, x_N) \dot{x}_{k} + 
\Gamma_i(x_1, \dots, x_N)=0, \ i=1, \dots, N.
\label{most-gen}
\eea
\noindent The $i^{th}$ particle of the system is subjected to a
velocity-independent force $-\Gamma_i$, gain/loss proportional
to $\eta_{ii}$ and a velocity mediated coupling $\eta_{ik} (i \neq k)$
with the $k^{th}$ particle. The particle is subjected to loss(gain) if
$\eta_{ii}$ is positive(negative) at a point in the configuration space.
The idea behind introducing an auxiliary system in Bateman's approach is
to make the system non-dissipative in the ambient space so that a Hamiltonian
formulation is possible. The system governed by Eq. (\ref{most-gen}) contains
both loss and gain, thereby, leading to the possibility that it may be
non-dissipative under certain condition and without the introduction of
an auxiliary system. This also raises the possibility of a Hamiltonian
formulation of the system governed by Eq. (\ref{most-gen}) without the
introduction of an ambient space or auxiliary system.

The condition for the system defined by Eq.
(\ref{most-gen}) to be non-dissipative may be determined following the
standard techniques in classical dynamics.
In particular, the equations of motion (\ref{most-gen}) can be
re-written as $2N$ coupled first order differential equations in terms of
two $2N$ dimensional vectors $\vec{\xi}$ and $\vec{G}(\xi_1, \dots,
\xi_{2N})$ as $\dot{\vec{\xi}}=\vec{G}$, where
\be
\xi_i \equiv x_i, \ \xi_{N+i} \equiv \dot{x}_i, \ G_i \equiv {\xi}_{N+i}, \ 
G_{N+i} \equiv -\Gamma_i(\xi_1, \dots, \xi_N)
-\sum_{k=1}^N \eta_{ik}(\xi_1, \dots, \xi_N) \xi_{N+k}.
\ee
\noindent The criteria for a non-dissipative system is that the flow preserves
volume in the $2N$ dimensional position-velocity state space spanned by
$\xi_i$'s, which is equivalent to putting the condition that $\vec{G}$ is
solenoidal, i.e. $\vec{\nabla}_{\xi} \cdot \vec{G}=0$. A straightforward
calculation shows,
\be
\vec{\nabla}_{\xi} \cdot \vec{G} \equiv 
\sum_{i=1}^{2N} \frac{\partial G_i}{\partial \xi_i}= \sum_{i=1}^N \eta_{ii},
\ee
\noindent implying that the condition for a non-dissipative system is
\be
 \sum_{i=1}^N \eta_{ii}=0.
\label{dissi-con}
\ee
\noindent The trivial solution $\eta_{ii}=0 \ \forall \ i$ is discarded from
the ambit of further discussions, since it does not correspond to gain and/or
loss for any individual particle. Eq. (\ref{dissi-con}) is also the condition
for the system governed by Eq. (\ref{most-gen}) to be identified as a system
with balanced loss and gain. The reason is that the flow preserves the volume
in the position-velocity state space spanned by $\xi_i$'s, although
individual particles are subjected to gain and/or loss.

One notable aspect of Bateman's formulation is that the dynamics of a
dissipative linear or nonlinear system is completely decoupled from that
of the dynamics of the corresponding auxiliary
system\cite{bateman,sagar,p6-deb}. The distinction between original system
and its auxiliary counterpart ceases to exists, if an interaction is added
to the ambient space Hamiltonian such that their dynamics are intertwined
to each other\cite{pkg-ds,ds-pkg1,p6-deb}. This feature may be used to rule out
the possibility that Eqs. (\ref{most-gen}) and (\ref{dissi-con}) do not
correspond to a combination of a hidden system plus its auxiliary part. In 
particular, the force $\Gamma_i$ on the $i^{th}$ particle may always be
chosen appropriately so that the dynamics of particles subjected to gain
and that of particles with loss are intertwined to each other. 
This gives rise to a new class of system with balanced loss and gain that
is different from the models obtained via Bateman's prescription.
Further discussions in this paper are restricted to this
distinct class of system with balanced loss and gain.

{\em Is it possible to find a Hamiltonian for the system with
balanced loss and gain that is governed by Eqs.  (\ref{most-gen}) and
(\ref{dissi-con}) without introducing any auxiliary system}? There is no
definite answer for the most general case with arbitrary $\Gamma_i$ and
$\eta_{ik}$. Even for a simpler case of $\eta_{ik}=(-1)^{i+1} \delta_{ik},
\Gamma_i=\frac{\partial \Gamma}{\partial x_i}$ with even $N$ and $\Gamma$
being the potential of rational Calogero model, no definite answer is
known for $N >2$ \cite{pkg-ds}. The Hamiltonian formulation of systems with
balanced loss and gain\cite{pkg-ds,p6-deb}, which is summarized below,
constitute a special case of Eq. (\ref{most-gen}), albeit encompassing a
very large class of such models. The importance of the formalism lies in the
fact that no auxiliary system is introduced. 

A Hamiltonian formulation of many-particle systems with space-dependent
balanced loss and gain is presented in Refs. \cite{pkg-ds,p6-deb}. The
analysis excludes constrained systems, systems with the dissipative term
depending nonlinearly on the velocity and any other non-standard Hamiltonian
formulations. The Hamiltonian is written as,
\bea
H=\Pi^T {\cal{M}} \Pi + V(x_1, x_2, \dots x_N),
\label{H}
\eea
\noindent where ${\cal{M}}$ is a $N \times N$ real symmetric matrix with
$X=(x_1, x_2, \dots x_N)^T$ and $\Pi=(\pi_1, \pi_2, \dots \pi_N)^T$
are $N$ coordinates and generalized momenta, respectively. The suffix $T$
in $O^T$ denotes the transpose of a matrix $O$. The matrix ${\cal{M}}$ may be
interpreted as a constant background metric. 
This is evident if the first term of $H$ is expressed in terms of the
generalized momenta $\pi_i$ as,
\bea
\Pi^T {\cal{M}} \Pi= \sum_{i,k=1}^N \pi_i {\cal{M}}_{ik} \pi_k.
\eea
The matrix ${\cal{M}}$ is non-singular and ${\cal{{M}}}^{-1}$ exists. However,
${\cal{M}}$ is not necessarily semi-positive definite. It may be noted in this
context that the Hamiltonian of the Bateman oscillator\cite{bateman} may also be
identified as defined in the background of a pseudo-Euclidean metric with the
signature $(1,-1)$ \cite{crv}. Moreover, all the Hamiltonian systems with
balanced loss and gain considered in the literature\cite{bpeng,ben,ben1,ivb,
sagar,ds-pkg,khare}, prior to the general formulation of such systems in
Ref. \cite{pkg-ds,ds-pkg1, p6-deb}, may be identified as defined in the
background of a pseudo-Euclidean metric.

The generalized momenta $\Pi$ is defined by,
\bea
\Pi=P+A F(X),
\label{gm}
\eea
\noindent where $P=(p_1, p_2, \dots, p_N)^T$  is the conjugate momentum
corresponding to the coordinate$X$, $F(X)=(F_1, F_2, \dots F_N)^T$ is $N$
dimensional column
matrix whose entries are functions of coordinates and $A$ is an $N \times N$
anti-symmetric matrix. The reason for considering $A$ to be anti-symmetric is
that the symmetric part of a general matrix replacing $A$ will not
contribute to the equations of motion resulting from $H$\cite{pkg-ds}.
Further, its contribution to the quantum mechanical wave-function can always
be removed via a gauge transformation.\cite{pkg-ds}. 
The generalized momenta $\Pi$, when written in component form, has the
following expression:
\bea
\pi_i = p_i + a_i, \ \ a_i \equiv \sum_{k=1}^N A_{ik} F_k,
\eea
where $a_i$ can be interpreted as components of a vector gauge potential
${\vec{a}}$. The introduction of the matrix $A$ to define the gauge potential
$\vec{a}$ instead of the vector field $F$ alone, is solely a matter of
convenience for formulating the problem in terms of arbitrary $F$.
In general, ${\vec{a}}$ consists of two parts, $\vec{a}=\vec{a}_r
+\vec{a}_f$. The vector $\vec{a}_r$ leads to realistic external magnetic
field. If there is no external magnetic field in the system, $\vec{a}_r$ may
be taken to be zero. The vector potential $\vec{a}_f$ is introduced to get the
effect of loss/gain terms. In fact, the magnitude of the analogous
magnetic field corresponding to $\vec{a}_f$ is identical with the loss/gain
co-efficients\cite{pkg-ds,p6-deb}. This interpretation is valid for Bateman
oscillator as well as all other examples considered so far. 

The Lagrangian corresponding to the
Hamiltonian $H$ in Eq. (\ref{H}) has the following form:
\be
{\cal{L}} = \frac{1}{4}\dot{X}^T {\cal{M}}^{-1} \dot{X} -\frac{1}{2} (\dot{X}^T
AF +F^T A^T \dot{X}) - V(x_1, x_2, \dots, x_N).
\label{lag}
\ee
\noindent The equations of motion derived from the Lagrangian (\ref{lag}) or
the Hamiltonian (\ref{H}) with the generalized momenta $\Pi$ defined by
Eq. (\ref{gm}) read,
\bea
\ddot{X}-2{\cal{M}}R\dot{X}+2{\cal{M}}\frac{\partial V}{\partial X}=0,
\label{X}
\eea
\noindent where the anti-symmetric matrix $R$ and 
$\frac{\partial V}{\partial X}$ are defined as follows:
\be
R \equiv AJ-(AJ)^T, \ \  
[J]_{ij} \equiv \frac{\partial F_i}{\partial x_j},\ \
\frac{\partial V}{\partial X}\equiv
\left(\frac{\partial V}{\partial x_i}, \frac{\partial V}{\partial x_2},
\dots \frac{\partial V}{\partial x_N}\right)^T.
\label{X-supp}
\ee
\noindent It may be noted that the anti-symmetric nature of $R$ follows from
its definition and is not imposed. Defining a matrix ${\cal{D}}$ as,
\be
{\cal{D}} \equiv {\cal{M}} R, 
\label{icon}
\ee
\noindent and comparing Eqs. (\ref{X}) with (\ref{most-gen}), one to one
correspondence may be made between these two equations through the following
identifications:
\be
\eta=-2 {\cal{D}}, \ \ \Gamma_i = 2 \sum_{k=1}^N M_{ik} \frac{\partial V}
{\partial x_k}.
\ee
\noindent It immediately follows from Eq. (\ref{dissi-con}) that the condition
for a balanced loss-gain system is $Tr({\cal{D}})=0$. The symmetric and
anti-symmetric nature of the matrices ${\cal{M}}$ and $R$, respectively,
ensures that the condition $Tr({\cal{D}})=\sum_{i=1}^N {\cal{M}}_{ik} R_{ki}
=0$ indeed holds for the Hamiltonian $H$. The Hamiltonian $H$ is identified as
describing a system with equally balanced loss and gain, since individual
particles are subjected to gain or loss such that net gain or loss of energy
is zero.

\subsection{Lorentz interaction \& a semi-positive definite ${\cal{M}}$}

In Refs.  \cite{pkg-ds,p6-deb}, it is shown on general ground that
for any symmetric matrix ${\cal{D}}$,
\be
\{{\cal{M}}, R\}=0, \ \{{\cal{M}}, {\cal{D}}\}=0, \ \{R, {\cal{D}}\}=0.
\label{anti}
\ee
\noindent This implies that the eigenvalues of the $2m \times 2m$ matrix
${\cal{M}}$ are $(\lambda_1, \dots, \lambda_m, -\lambda_1, \dots, -\lambda_m)$,
while for odd-dimensional ${\cal{M}}$ an extra eigenvalue $1$ is added to it.
Thus, the matrix ${\cal{M}}$ is not semi-positive definite for a symmetric
${\cal{D}}$. Stable classical solutions are possible for specific choices of
the potentials and parameter ranges. However, except for one known case
involving a quasi-exactly solvable model with anaharmonic 
interaction\cite{p6-deb}, quantum bound states exist
for such systems only if the eigenvalue problem is extended to complex
domain. In particular, normalizable eigenfunctions for bound states
are obtained in specific Stoke wedges on the complex plane. Although
the complex extension of quantum mechanics is mathematically consistent,
no experimental verification of this has been found so far.
One of the motivations of this article is to present a model independent
formulation of Hamiltonian system with balanced loss and gain so that
the quantum problem is well-defined on the real line. The non-normalizability
of eigenfunctions on the real line may be traced back to the non-positivity
of ${\cal{M}}$ which is a consequence of a symmetric ${\cal{D}}$ through
the relations in Eq. (\ref{anti}).

Investigations\cite{pkg-ds,p6-deb,ds-pkg1} on Hamiltonian systems with
balanced loss and gain have been so far restricted to a diagonal ${\cal{D}}$
which is a symmetric matrix. This is primarily because only the diagonal
elements of ${\cal{D}}$ are relevant for determining whether a system is
dissipative or non-dissipative. Nevertheless, the off-diagonal elements of
${\cal{D}}$ give rise to a variety of interesting physical effects. For
example, the equation of motion for a particle subjected to Lorentz force
in appropriate units is given by,
\be
\ddot{x}_i = E_i + \epsilon_{ijk} \dot{x}_j B_k, \ i=1, 2, 3,
\label{lorentz}
\ee
\noindent where $E_i\equiv E_i(x_1,x_2, x_3)$ and $B_i\equiv B_i(x_1,x_2,x_3)$
are components of electric and magnetic fields. Eq. (\ref{lorentz}), when cast
into the form of Eq. (\ref{X}), leads to an anti-symmetric ${\cal{D}}$ with its
elements ${\cal{D}}_{ij}=-\frac{1}{2} \epsilon_{ijk} B_k$. Further, when the
equations governing synchronization of velocity-coupled systems are cast
into the form (\ref{X}), the matrix ${\cal{D}}$ contains diagonal as well as
off-diagonal elements. For example, the non-vanishing diagonal as well as
off-diagonal elements of ${\cal{D}}$ appear in the study of stochastic
synchronization of oscillation in systems of velocity-coupled oscillators with
individual chaotic dynamics\cite{qe} for which off-diagonal part of
${\cal{D}}$ is symmetric. Similarly, in the case of synchronization of
velocity-coupled limit-cycle oscillators, the off-diagonal part of ${\cal{D}}$
is symmetric\cite{qe1,qe2}. The trace of ${\cal{D}}$ is non-vanishing for both
the cases and the systems are dissipative. Similar velocity dependent coupling
also arises in  the description of partially ionized plasma\cite{pande}.
Thus, the domain of applicability of Hamiltonian systems with balanced loss
and gain may be enlarged by considering off-diagonal elements of ${\cal{D}}$.

A symmetric ${\cal{D}}$ leads to negative eigenvalues of ${\cal{M}}$.
Although ${\cal{M}}$ may be chosen to be semi-positive definite for an
anti-symmetric ${\cal{D}}$, no gain and/or loss term can be incorporated in
the system due to ${\cal{D}}_{ii}=0, \forall \  i$. Thus, the matrix
${\cal{D}}$ can neither be symmetric nor anti-symmetric in order to
describe a Hamiltonian system with balanced loss and gain such that 
the symmetric matrix ${\cal{M}}$ is semi-positive definite.
In general, ${\cal{D}}$ may always be decomposed as the sum of a
diagonal matrix $D$, a symmetric matrix with vanishing diagonal elements
$D_O$ and an antisymmetric matrix ${\cal{D}}_A$:
\be
{\cal{D}}= D + D_O + {\cal{D}}_A.
\label{lorai}
\ee
\noindent It may be noted that ${\cal{D}}_S=D+D_O$ is a symmetric matrix
and $Tr({\cal{D}})=0$ implies $Tr(D)=0$. The elements of $D$ are related
to the loss/gain co-efficients, while elements of ${\cal{D}}_A$ are related
to the Lorentz interaction. The elements of $D_O$ give rise to coupling among
different particles through their velocities, which is different from Lorentz
interaction and appear in the study of many systems including synchronization
of different types of oscillators\cite{qe1,qe2} and in the description of
partially ionized plasma\cite{pande}. All three matrices correspond
to interesting physical situations. It may be possible to describe a
Hamiltonian system with balanced loss and gain so that ${\cal{M}}$ is
positive definite, even if $D_O$ is taken to be a zero matrix. However,
this is impossible if either $D$ or ${\cal{D}}_A$ is taken to be a zero matrix.
One very important
consequence of a non-vanishing ${\cal{D}}_A$ is that balancing of gain/loss
terms does not necessarily occur in a pair-wise fashion, unlike the previous
cases\cite{pkg-ds, ds-pkg1,p6-deb}. The balancing of loss/gain terms are
possibles in as many ways as the solutions of $Tr({\cal{D}})=0$ can be
realized. For example, $N_g$ particles may be subjected to gain with
coefficients $g_i$ and $N_l$ particles may be be subjected to loss with
coefficients $l_i$, such that $\sum_{i=1}^{N_g} g_i - \sum_{i=1}^{N_l} l_i=0,
N=N_g+N_l$. The case of pair-wise balancing of loss/gain terms appears as a
special case. Such a formulation allows more flexibility in constructing and
controlling Hamiltonian systems with balanced loss and gain. 

\subsection{Representation of matrices}

The matrix ${\cal{D}}$ can be decomposed as the product of a symmetric and
an anti-symmetric matrix if it is similar to $-{\cal{D}}$\cite{LR}. However,
the representation of the matrices ${\cal{M}}$, $R$ and ${\cal{D}}$ satisfying
Eq. (\ref{icon}) is not unique and various choices may be made depending on
physical situations. It may be noted that a diagonal ${\cal{M}}$ necessarily
leads to a ${\cal{D}}$ with all of its diagonal elements to be zero
and can not describe a system with balanced loss and gain. On the
other hand, a traceless ${\cal{M}}$ is not semi-positive definite. Thus,
the necessary conditions for the choice of ${\cal{M}}$ is the following:
\begin{enumerate}
\item $Trace({\cal{M}}) \neq 0$
\item $[{\cal{M}}]_{ij} \neq 0, \ i \neq j$ for one or more pairs of $(i,j)$
\end{enumerate}
\noindent The sufficient condition for ${\cal{M}}$ to be positive-definite
is to be checked separately. The product of any positive definite ${\cal{M}}$
with non-vanishing off-diagonal elements and an anti-symmetric matrix
${{R}}$ will determine the matrix ${\cal{D}}$. The matrix ${\cal{D}}$
can always be decomposed as in Eq. (\ref{lorai}):
\be
{\cal{D}}_S=\frac{1}{2} \left [ {\cal{M}}, R \right ], \ \
{\cal{D}}_A=\frac{1}{2} \left \{{\cal{M}}, R \right \}.
\ee 
\noindent A few representations of the matrices are presented in
Appendix-A for arbitrary $N$, where the balancing of loss-gain terms does
not necessarily occurs in a pair-wise fashion. If the off-diagonal elements
of ${\cal{D}}$ are to be related to contributions coming from Lorentz
interaction only, then ${\cal{D}}_O$ must be taken to be zero and ${\cal{D}}$
is decomposed as, 
\be
{\cal{D}} \equiv D + {\cal{D}}_A.
\label{f-lorai}
\ee  
\noindent A particular representation of Eqs. (\ref{icon}) and (\ref{f-lorai})
for $N=2m$ with pair-wise balancing of loss/gain terms is considered
here to show that ${\cal{M}}$ can indeed be
chosen to be positive definite. The matrix ${\cal{M}}$ is chosen as,
\be
{\cal{M}} = M + \alpha^2 I_{2m},
\label{mnew}
\ee
\noindent where $M$ is a traceless $2m \times 2m$ symmetric matrix,
$I_{2m}$ is the $2m \times 2m$ identity matrix and $\alpha$ is a real
parameter. Note that both ${\cal{M}}$ and $M$ are simultaneously
diagonalizable. Substituting ${\cal{M}}$ in Eq. (\ref{icon}), the decomposition
of ${\cal{D}}$ as in Eq. (\ref{lorai}) is obtained with the identification,
\be
D=M R, \ {\cal{D}}_A = \alpha^2 R.
\label{old}
\ee
\noindent The first equation of Eq. (\ref{old}) involving symmetric matrices
$M, D$ and anti-symmetric matrix $R$ implies that they anti-commute with
each other. Thus, the eigenvalues of ${\cal{M}}$ are $\alpha^2 \pm \lambda_i,
i=1, \dots m$. The condition for semi-positive definite ${\cal{M}}$ is,
$\alpha^2 \geq \max_i \lambda_i$. Matrix representation of the first equation
of Eq. (\ref{old}) is suffice to completely specify the representation
of ${\cal{M}}, {\cal{D}}$ and $R$. A few representations of the equation
$D=M R$ have been discussed in Ref. \cite{pkg-ds} for systems with constant
balanced loss-gain terms. For the case of space-dependent balanced loss-gain
terms with $F_i\equiv F_i(x_{2i-1},x_{2i})$, a particular representation of
the matrices is presented in Ref. \cite{p6-deb}, which is reproduced below
for further discussions:
\bea 
M=I_m\otimes\sigma_x,\ \ \  \ A=\frac{-i\gamma}{2}I_m\otimes\sigma_y,\ \
D=\gamma \chi_m\otimes \sigma_z, \ \ [\chi_m]_{ij}=\frac{1}{2}\delta_{ij}
Q_i(x_{2i-1}, x_{2i}),
\label{repre}
\eea
\noindent where $\sigma_x, \sigma_y, \sigma_z$ are Pauli matrices and $I_m$ is
$m \times m$ identity matrix. The $m$ functions $Q_i$ appearing in the
$m \times m$ diagonal matrix $\chi_m$ is determined as,
\be
Q_a(x_{2a-1},x_{2a}) = Trace (V_a^{(2)}), \ \
V_a^{(2)} \equiv
\bp
{\frac{\partial F_{2a-1}}{\partial x_{2a-1}}}
& {\frac{\partial F_{2a-1}}{\partial x_{2a}}}\\
{\frac{\partial F_{2a}}{\partial x_{2a-1}}} &
{\frac{\partial F_{2a}}{\partial x_{2a}}}
\ep.
\label{repq}
\ee
\noindent It may be noted that the matrix $J$ has a block-diagonal form
for the choices of $F_i\equiv F_i(X_{2i-1},x_{2i})$ and $R$ may be determined
from the first two equations of Eq. (\ref{X-supp}) by using the expressions
of $A$ and $J$:
\be
R = \frac{\gamma}{2} \sum_{i=1}^m U_i^{(m)} \otimes
\bp
0 && - Q_i(x_{2i-1},x_{2i})\\
Q_i(x_{2i-1},x_{2i}) && 0
\ep
, 
\ \ [U_a^{(m)}]_{ij} = \delta_{ia} \delta_{ja},
\label{repare}
\ee
\noindent where $U_a^{(m)}$ are $m$ numbers of $m \times m$ matrices.
The eigenvalues of $M$ are $\pm 1$. Thus, ${\cal{M}}$ is positive-definite for
$\alpha^2 > 1$. This
completely specifies the representation of ${\cal{M}}$, $R$ and ${\cal{D}}$.
This representation for even number of particles can be generalized\cite{pkg-ds}
to $N=2m+1$ such that the dynamics of $x_{2m+1}$ does not contain any gain/loss
term and interacts with all other particles through the interaction potential.
This can be achieved by adding an extra column and a row to ${\cal{M}}, R,
{\cal{D}}$ with all vanishing elements except for ${\cal{M}}_{2m+1,2m+1}$. 
The case of constant balanced loss and gain may be reproduced\cite{pkg-ds}
for $F_{2i-1}=x_{2i-1}$ and $F_{2i}=x_{2i}$. 

The representation specified by Eqs. (\ref{repre}, \ref{repq}, \ref{repare})
determines ${\cal{D}}$ with ${\cal{D}}_O$ being a null matrix. However, the
same representations with a modified ${\cal{M}}$,
\be
{\cal{M}} = \beta_1 M + \alpha^2 I_{2m} + \beta_2 I_m \otimes \sigma_z,
\beta_1, \beta_2 \in \Re,
\ee 
\noindent may be used to get the following expression of 
${\cal{D}}= \beta_1 D + \beta_2 D_O + \alpha^2 R$ with $D_O$ having
non-vanishing elements:
\be
D_O = \frac{\beta_2 \gamma}{2} \sum_{i=1}^m U_i^{(m)} \otimes
\bp
0 && - Q_i(x_{2i-1},x_{2i})\\
- Q_i(x_{2i-1},x_{2i}) && 0
\ep.
\ee
\noindent The eigenvalues of ${\cal{M}}$ are $\alpha^2 \pm \sqrt{\beta_1^2
+\beta_2^2}$ with multiplicity $m$ for each eigenvalue. The matrix ${\cal{M}}$
is positive definite for $\alpha^2 > \sqrt{\beta_1^2+\beta_2^2}$. 

\subsection{Hiding the loss-gain terms}

It is known\cite{pkg-ds,ds-pkg1,p6-deb} that Hamiltonian systems with
balanced loss-gain and without any external magnetic field can always be
interpreted as defined in the background of a pseudo-Euclidean metric and
subjected to external `analogous magnetic field' having the same spatial
form as the gain-loss co-efficient. The equations of motion in the new
co-ordinate system do not contain any gain-loss terms. A similar investigation
for the case of a realistic external magnetic field is performed in this
section. The matrix ${\cal{M}}$ being a real symmetric matrix, it can be
diagonalized by an orthogonal matrix $\hat{O}$, i.e. $M_d= \hat{O}^T {\cal{M}}
\hat{O}$. The matrix $M_d$ and two other matrices $S$ and $\eta^a$ are defined
as follows:
\be
[M_d]_{ij}=\epsilon_i \delta_{ij} {\lvert \lambda_i \rvert}, \ \
[S]_{ij}=\delta_{ij} \sqrt{{\lvert \lambda_i \rvert}}, \ \
[\eta^a]_{ij}= \delta_{ij} \epsilon_i,
\label{eta}
\ee
\noindent where $\lambda_i$'s are the eigenvalues of the matrix ${\cal{M}}$
and $\epsilon_i$'s take value of $1$ or $-1$. It may be noted that
$\epsilon_i$'s keep track of whether a particular eigenvalue $\lambda_i$ is
positive or negative. The parameter space of the system can be divided into
at most $N+1$ regions depending on the number of positive eigenvalues
of the matrix ${\cal{M}}$. For example, the region with $N$ positive
eigenvalues may be denoted as Region-I, the region with $N-1$ positive
eigenvalues may be denoted as Region-II and so on with the region with
no positive eigenvalues is denoted as Region-$(N+1)$. The superscript of
$\eta^a$ labels each region and for a fixed `$a$', $\eta^a$ is valid in
that particular region only. For example, $\eta^I=I_N$  in Region-I, while
$\eta^{N+1}=-I_N$ in Region-$(N+1)$. The matrix $\eta^a$ is to be interpreted
as the background metric for an effective description of the system defined by
the Hamiltonian $H$ and equations of motion following from it in Eqs.
({\ref{H}},{\ref{X}}).

The system described by the  equations of motion (\ref{X}) is non-dissipative,
although individual particles are subjected to gain or loss. Are the appearance
of gain-loss terms an artifact of choice of the co-ordinate system and can be
removed completely in an another frame? In order to answer
this question, a new co-ordinate system\cite{pkg-ds} is defined as follows:
\be
{\cal{X}}= S^{-1} \hat{O}^T X, \ \ {\cal{P}}= S \hat{O}^T P, \ \
\label{trans-cor}
\ee 
\noindent where ${\cal{X}}\equiv ({\cal{X}}_1, {\cal{X}}_2, \dots,
{\cal{X}}_N)^T$ and ${\cal{P}}\equiv ({\cal{P}}_1, {\cal{P}}_2, \dots,
{\cal{P}}_N)^T$.
The matrix $S$ along with its inverse $S^{-1}$ is used to generate canonical
scale transformation for the rotated co-ordinates $\hat{X}\equiv \hat{O}^T X$
and the rotated momenta $\hat{P} \equiv \hat{O}^T P$, where $\hat{X} \equiv
(\hat{X}_1, \hat{X}_2, \dots \hat{X}_N)^T$ and $\hat{P} \equiv (\hat{P}_1,
\hat{P}_2, \dots \hat{P}_N)^T$. The kinetic energy term of $H$ is in diagonal
form in the rotated co-ordinate system $\hat{X}, \hat{P}$. The scale
transformation is performed so that all the coefficients of the
$\hat{P}_i^2$ terms are normalized to unity. Thus, the background metric in
the co-ordinate system by $\cal{X}$ may be identified as (pseudo-)Euclidean
one. The matrices ${\cal{M}}$, $R$ and ${\cal{D}}$ are transformed as follows:
\bea
&& S^{-1} \hat{O}^T {\cal{M}} \hat{O} S^{-1}=\eta^a,\nonumber \\
&& [{\cal{R}}]_{ij} = [O^T R O]_{ij} \sqrt{{\lvert \lambda_i \rvert}
{\lvert \lambda_j \rvert}},\nonumber \\
&& S^{-1} O^T {\cal{D}} O S = \eta^a {\cal{R}},
\label{simi-mat}
\eea
\noindent which are obtained by simply using the rules of matrix
multiplication. It may be noted that ${\cal{R}}$ is anti-symmetric, since
$R^T=-R$. The first and the third equations in Eq. (\ref{simi-mat}) are
different in different regions, while the second equation has the same
expression in all the regions. It may be noted that $S^{-1}$ and $S$
are not generating any similarity transformation for $\hat{O}^T {\cal{M}} O$
and $\hat{O}^T R \hat{O}$, respectively. Consequently, eigenvalues of
${\cal{M}}$ and ${{R}}$ are modified. However, $OS$ indeed generates a
similarity transformation for ${\cal{D}}$ and the eigenvalues of
${\cal{D}}$ remain unchanged.

The generalized momenta $\Pi$ transforms like $P$ and denoted as $\hat{\Pi}$
in the new co-ordinates:
\be
\hat{\Pi} \equiv S\hat{O}^T \Pi= {\cal{P}} + \frac{\cal{R}}{2} {\cal{X}},
\label{pini}
\ee
\noindent which may be obtained by using the canonical transformation
in Eq. (\ref{trans-cor}). The Hamiltonian and the equations of motion
resulting from it in different regions have the forms: 
\bea
&&H^a={\hat{\Pi}}^T \eta^a {\hat{\Pi}} + {\cal{V}}({\cal{X}}_1, {\cal{X}}_2,
\dots, {\cal{X}}_N),\nonumber \\
&& \ddot{\cal{X}} -  2 \eta^a {\cal{R}} \dot{\cal{X}} +
2 \eta^a \left ( \frac{\partial {\cal{V}}}{\partial {\cal{X}}} \right )=0,
\ \ {\cal{V}}({\cal{X}}_1, {\cal{X}}_2, \dots, {\cal{X}}_N)= V(x_1, x_2,
\dots, x_N). 
\label{hami-eq}
\eea
\noindent There are no loss and/or gain terms in the equations of motion
(\ref{hami-eq}) in terms of ${\cal{X}}$, since $(\eta^a {\cal{R}})_{ii}=0
\ \forall \ i, \ a$. The effect of removing the gain-loss terms from Eq.
(\ref{X}) is to modify the coefficients of the velocity-mediated non-Lorentzian
interaction and/or the magnitude of the magnetic field, since
${\cal{D}}_{ij} \neq (\eta^a R)_{ij}, \ \forall \ i, j$. In particular, 
the velocity dependent non-Lorentzian interaction vanishes in region-I and
region-(N+1) in the effective description, since $\eta^a R$ becomes
anti-symmetric. The applied magnetic field is modified to an effective
magnetic field whose components are related to the elements of the
$\eta^a {\cal{R}}$. In Region-II to Region-$N$, both non-Lorentzian and
Lorentzian interactions are present with modified coupling co-efficients,
since $\eta^a {\cal{R}}$ is neither symmetric nor anti-symmetric.
However, with specific representations of ${\cal{M}}$ and $R$, $\eta^a R$ may
be made to be symmetric or anti-symmetric.

A few advantages of using Eq. (30) over Eqs. (\ref{H}) and (\ref{X}) in
analyzing the system may be mentioned. In Region-I and Region-(N+1), the
quantum problem may be defined on the real line instead of Stokes wedges
on the complex plane. There is a vast literature on systems in presence of
external magnetic field, which may be used efficiently to study the classical
as well as quantum system. In Region-II to Region-N, use of imaginary scaling
of the co-ordinates\cite{ali-review} associated with negative signatures may
be helpful to
analyze the quantum Hamiltonian. As far as the classical system is concerned,
which form is to be used for finding the solution is a matter of convenience.
However, it seems by analyzing a number of problems that Eq. (30) is
relatively simpler to solve than Eq. (\ref{X}).

\section{Landau Hamiltonian with balanced loss and gain}

It has been shown that that an external magnetic field allows the matrix
${\cal{M}}$ to be positive-definite and thereby, raising the possibility of
improved stability properties of the system. A system subjected to 
(non-)uniform magnetic field appears in many areas of physics. The examples
include Zeeman effect, Cyclotron, Hall effect, spintronics, neutron stars,
plasma, etc. Thus, it is pertinent to consider systems where magnetic field is
essentially required to describe a physical phenomenon and generalize them
by including balanced loss-gain terms. The celebrated Landau Hamiltonian 
with balanced loss-gain is considered in this section at the classical as
well quantum level. 

\subsection{Classical system}

A two dimensional system with uniform balanced loss-gain terms is considered.
Thus, the functions $F_{i}$ are chosen as $F_i=x_i$ and $N=2$. The condition
of constant loss-gain terms imposed by the choice of $F_i$ also implies that
the matrix $J$ is an identity matrix and hence, $A=\frac{1}{2} R$.
The representation of the matrices ${\cal{M}}, R,
{\cal{D}}$ may be considered as follows:
\be
{\cal{M}}= \frac{1}{2}
\bp
{B+C} && \gamma\\ \gamma && B-C
\ep,
R =
\bp
{0} && 1\\ -1 && 0
\ep,
{\cal{D}}= \frac{1}{2}
\bp
{-\gamma} && B+C\\ -(B-C) && \gamma .
\ep
\label{landau-rep}
\ee
\noindent Decomposing ${\cal{D}}$ as in Eq. (\ref{lorai}), 
it is apparent from Eq. (\ref{X}) that the particle is subjected to a uniform
magnetic field $B$ along the perpendicular to the `$x_1-x_2$'-plane. A change
in the direction of the magnetic field is accomplished by taking
$B \rightarrow -B$. The vanishing field $B=0$ corresponds to
$Trace({\cal{M}})=0$ and ${\cal{M}}$ is not positive-definite. Apart from the
Lorentz interaction\footnote{An appropriate choice of $V$ produces an external
electric field.}, a velocity mediated coupling between the two degrees of
freedom with the strength $C$ is also present in the system. 
The discussions in this article will be based on generic $B$ and $C$ unless
specified otherwise.
The case of pure Lorentz interaction may be discussed by employing
the limit $C=0$. The particle is subjected to balanced loss and gain
with the strength $\gamma$. The eigenvalues of the matrix ${\cal{M}}$ are,
\be
\lambda_{\pm} = \frac{1}{2} \left ( B \pm \bigtriangleup \right ), \ \
\bigtriangleup \equiv \sqrt{ C^2 + \gamma^2}.
\ee
\noindent 
The matrix ${\cal{M}}$ becomes singular for $B=\pm \bigtriangleup$ and these
two conditions determine the boundaries of three different regions in 
the parameter space of $B$ and $\bigtriangleup$:
\begin{itemize}
\item Region-I ( {\bf $B > \bigtriangleup$} ): The matrix ${\cal{M}}$ is 
positive-definite. This condition for the case of pure Lorentz
interaction(i.e. $C=0$) implies that the magnitude of the
external magnetic field must be greater than the magnitude of the
gain/loss co-efficient $\gamma$. It appears that a positive-definite
${\cal{M}}$ has not been considered earlier in the literature. Thus, this
region is of special interest for the present article.
The diagonal matrix $\eta^I=I_2$.

\item Region-II ( {\bf $ -\bigtriangleup < B < \bigtriangleup$}):
One of the eigenvalues $\lambda_-$ is negative, while $\lambda_+$ is
positive. All previous studies\cite{ben, pkg-ds, ds-pkg1, p6-deb}
on Hamiltonian systems with balanced loss/gain dealt with the case
$\lambda_+=-\lambda_-$, which is contained in the present case for $B=0$.
However, for $B \neq 0$, $\lambda_+$ and $-\lambda_-$ are different.
It is expected that for $B \neq 0$ there may exist sub-regions within
this region in which the classical system admits periodic solution and
the corresponding quantum theory admits well defined bound states in
specific Stoke wedges as in the case for $B=0$. The diagonal matrix
$\eta^{II}=\sigma_z$

\item Region-III ( {\bf $ B < - \bigtriangleup$}): Both the eigenvalues
$\lambda_{\pm}$are negative and it appears that such a situation has not
been considered earlier in the investigations on Hamiltonian system with
balanced loss/gain. The Hamiltonian is not bounded from below, but, bounded
from above. An inclusion of an appropriate potential $V$ in the system may
allow the Hamiltonian to be bounded from below, thereby, raising the
possibility of classical as well as quantum bound states. The diagonal matrix
$\eta^{III}=-I_2$.
\end{itemize}

The orthogonal matrix $\hat{O}$ that diagonalizes ${\cal{M}}$ to
$M_d= \hat{O}^T {\cal{M}} \hat{O}$ has the form:
\bea
M_d =
\bp
\lambda_+ && 0\\
0 && \lambda_-
\ep, \ \ \
\hat{O} = \frac{1}{\sqrt{\gamma^2+(\bigtriangleup-C)^2}}
\bp
 {\gamma} && -(\bigtriangleup-C)\\
{\bigtriangleup-C} && {\gamma}
\ep.
\eea
\noindent The matrix ${\cal{M}}$ is diagonal for $\gamma=0$ for which
$\hat{O}$ is not defined. The diagonal matrix $S$ has the non-vanishing
elements, $[S]_{11} = \sqrt{{\lvert \lambda_+ \rvert}}$ and
$[S]_{22}=\sqrt{{\lvert \lambda_- \rvert}}$. The transformations
(\ref{trans-cor}), with $\hat{O}$ and $S$ as given above, consist of a
rotation in the `$x_1-x_2$'-plane by an angle $\theta= \tan^{-1} (
\frac{\bigtriangleup -C}{\gamma})$ followed by scaling of $\hat{X}_1$ and
$\hat{X}_2$ by $\frac{1}{\sqrt{\lvert \lambda_+ \rvert}}$  and
$\frac{1}{\sqrt{\lvert \lambda_-\rvert }}$, respectively.
The canonical transformation is not defined at the two boundaries ($B=\pm
\bigtriangleup$) of the three regions and within each region for $\gamma =0$
and $C \neq 0$. Thus, the limit $\gamma=0$ is singular for $C \neq 0$.
The results for standard Landau Hamiltonian can be reproduced from the results
in the transformed co-ordinates $({\cal{X}}_1, {\cal{X}}_2)$ by taking $C=0$
and then imposing the limit $\gamma \rightarrow 0$. The angle
$\theta=\frac{\pi}{4}$ for $C=0, \gamma \neq 0$, which corresponds to
velocity-dependent force due to Lorentz interaction only.

The generalized $\hat{\Pi}$ in the new co-ordinates has the form:
\be
\hat{\Pi} = {\cal{P}} + \frac{\lvert \omega \rvert}{4} R {\cal{X}},
\ \ \omega \equiv \sqrt{B^2-\bigtriangleup^2}.
\label{pini}
\ee
\noindent The Hamiltonian and the equations of motion in this new co-ordinate
system have the following expressions:
\be
H^a={\hat{\Pi}}^T \eta^a {\hat{\Pi}}, \ \ 
\ddot{\cal{X}} -  {\lvert \omega \rvert} \eta^a {{R}} \dot{\cal{X}}=0.
\label{fini}
\ee
\noindent It may be noted that there are no gain-loss terms in this new
co-ordinate system, since the diagonal elements of $\eta^a R$ are zero.
The absence of the gain-loss terms is compensated by the appearance of
an effective magnetic field with its magnitude receiving contributions
from the realistic external magnetic field and an analogous magnetic field.
The Hamiltonian and the equations of motion in Eq. (\ref{fini})
correspond to the standard Landau problem for $\eta^a=I_2$ for which
the centre of the cyclotron motion is a constant of motion. The same analysis
may be generalized in a straightforward way to other allowed forms of $\eta^a$
to find corresponding constants of motion. The components of the vector
$C^a=(C_1^a, C_2^a)^T$ are two constants of motion of the system:
\be
C^a= {{\cal{X}}} + \frac{1}{\lvert \omega \rvert} R \eta^a \dot{\cal{X}}=
\hat{O} S \left ( {X} + \frac{1}{\lvert \omega \rvert} R \eta^a \dot{X}
\right ).
\ee
\noindent The first expression on the right
side of $C^a$ is the centre of cyclotron motion for the system defined by Eq.
(\ref{fini}), which is re-written in terms of the original variables $X$ in
the second expression by using Eq. (\ref{trans-cor}). It may be checked that
$\frac{d C^a}{dt}=0$ by using the equations of motion and the identities
$R^2=-I, (\eta^a)^2=I, \ \forall \ a$.
It may be noted that two independent constants of motion may
be chosen for the system, depending on the physical requirements, by 
taking appropriate combinations of $C_1^a$ and $C_2^a$. The reason for the
particular choice of $C^a$ is that it may be identified as the center of the
cyclotron motion for $\eta^a=I_2$. Two complex parameters $\xi_1, \xi_2$ along
with their polar decompositions are introduced as follows,
\bea
&& \xi_1= \frac{2}{\sqrt{\lvert \omega \rvert}} \left (
\sqrt{\lvert \lambda_+ \rvert} \cos \theta + 
i \sqrt{\lvert \lambda_- \rvert} \ \sin \theta \right )={\lvert \xi_1
\rvert} e^{i \phi_1},\nonumber \\ 
&& \xi_2= \frac{2}{\sqrt{\lvert \omega \rvert}} \left (
-\sqrt{\lvert \lambda_+ \rvert} \sin \theta +
i \sqrt{\lvert \lambda_- \rvert} \cos \theta \right )={\lvert \xi_2
\rvert} e^{i (\phi_2+\frac{\pi}{2})},\nonumber \\ 
&& \phi_1= \tan^{-1} \left ( \sqrt{\frac{\lvert \lambda_- \rvert}
{\lvert \lambda_+ \rvert} } \ \tan \theta \right ),\ \
\phi_2= \tan^{-1} \left ( \sqrt{\frac{\lvert \lambda_+ \rvert}
{\lvert \lambda_- \rvert} } \ \tan \theta \right ),
\eea
\noindent which will be used for presenting the results in terms of the
original co-ordinates $(x_1, x_2)$ in a compact form. It may be recalled
that $C=0$ corresponds to the situation with Lorentz interaction only for
which $\theta=\frac{\pi}{4}$ and $\xi_1=-\xi_2^*$. The nature of the
solutions in three different regions are different and is described
below separately. It is worth emphasizing here that no solutions  
at the two boundaries $B=\pm \bigtriangleup$ separating the three regions
are presented. It may be noted at this point that the Hamiltonian formulation
of balanced loss-gain system is based on the assumption of a non-singular
${\cal{M}}$, which is violated at these two boundaries. Thus, the discussions
on solutions at the boundaries are beyond the purview of this article.

\subsubsection{Region-I} 

The solutions may be written as,
\bea
&& x_1= C_1^I + \frac{\lvert \xi_1 \rvert}{2 {\lvert \omega \rvert}}
{\lvert Z \rvert} \
\cos ({\lvert \omega \rvert}t - \phi_1),\nonumber \\
&& x_2= C_2^I - \frac{\lvert \xi_2 \rvert}{2 {\lvert \omega \rvert}}
{\lvert Z \rvert} \
\sin ({\lvert \omega \rvert} t - \phi_2), \ \
Z\equiv \xi_1 \dot{x}_2 + \xi_2 \dot{x}_1.
\label{sol-classi}
\eea
\noindent The quantity $\frac{\lvert Z \rvert}{\lvert \omega \rvert}$ is
related to the cyclotron radius in the transformed co-ordinate $({\cal{X}}_1,
{\cal{X}}_2)$. The Hamiltonian $H$ in Region-I describes a system with
balanced loss and gain for which its kinetic energy term is semi-positive
definite. The system is interpreted as that of a particle moving on an
Euclidean plane and subjected to an effective external magnetic field
${\lvert \omega \rvert}$, which is less than the applied magnetic field $B$.
The role of the external magnetic field with a lower bound on its magnitude
$B > \bigtriangleup$ is essential to achieve this. The effect of the inclusion
of balanced loss-gain to the Landau Hamiltonian is to have a reduced value of
the cyclotron frequency ${\mid \omega \mid}$ compared to its original magnitude
$B$. The co-ordinate transformation induces a similarity transformation on the
matrix ${\cal{D}}$, keeping its purely imaginary eigenvalues $ \pm i \omega$
unchanged, the modulus of which determines the cyclotron frequency. Thus, the
motion described in either of the co-ordinate systems has the same cyclotron
frequency. However, the trajectories of the particle are circle in the
co-ordinates $( {\cal{X}}_1, {\cal{X}}_2)$, while it is ellipse in the
co-ordinates $(x_1, x_2)$. The unequal scaling of the two co-ordinates $(x_1,
x_2)$ allows the elliptic orbits to be viewed as circular orbits in the
co-ordinate system $( {\cal{X}}_1, {\cal{X}}_2)$. In general, for
$C \neq 0, \gamma \neq 0$, the reduced value of the cyclotron frequency
and an elliptic orbit is due to the effect of both the gain-loss coefficient
$\gamma$ and the velocity mediated coupling $C$. However, this result is valid
even if only Lorentz interaction is considered and the velocity mediated
coupling $C=0$. In particular, in the limit of pure Lorentz interaction($C=0$),
the co-efficient of the time-varying part for the two solutions are identical,
since ${\lvert \xi_1 \rvert}= {\lvert \xi_2 \rvert}$ for
$\theta=\frac{\pi}{4}$. However, the phases $\phi_1$ and $\phi_2$ are
different.

\subsubsection{Region-II} 
The solutions may be written as,
\be
X=C^{II} + \frac{1}{2 {\lvert \omega \rvert}} \left ( Z^2 +{Z^*}^2 \right )^
{\frac{1}{2}} \hat{O} S 
\bp
\cosh {\lvert \omega \rvert} t\\
\sinh {\lvert \omega \rvert} t
\ep.
\label{2sol-classi}
\ee
The kinetic energy term is not semi-positive definite. The Hamiltonian may be
interpreted as that of a particle moving in the background of a
pseudo-Euclidean metric that is subjected to an effective external magnetic
field ${\lvert \omega \rvert}$. The magnitude of the external magnetic field
is less than the magnitude of the analogous magnetic field. The known results
for generic Hamiltonian of this form\cite{pkg-ds,ds-pkg1,p6-deb} for $B=C=0$
appears as a special case in this region with an `analogous magnetic field'
having the magnitude $\gamma$. For $B=0, C \neq 0$, the same interpretation is
valid with the magnitude of the `analogous magnetic field' being
$\bigtriangleup$. However, the description changes for $B \neq 0$ for which
the effective magnetic field receives contribution from both the applied as
well as analogous magnetic field. The Hamiltonian $H^{II}$ does not admit any
periodic solution, as is evident from Eq. (\ref{2sol-classi}). The quantity
$Z^2+{Z^*}^2$ is fixed by initial conditions and may be
chosen to be positive so that $(Z^2+{Z^*}^2)^{\frac{1}{2}}$ is real. The
solutions diverge in the limit of large $t$. The Hamiltonian for $B \neq0,
C\neq 0$ may admit periodic solutions if an appropriate non-vanishing
${\cal{V}}$ is added to the system. Such an investigation is beyond the scope
of the present article.

\subsubsection{Region-III}

The solutions are,
\bea
&& x_1= C_1^{III} + \frac{\lvert \xi_1 \rvert}{2 {\lvert \omega \rvert}}
{\lvert Z \rvert} \
\cos ({\lvert \omega \rvert} t - \phi_1),\nonumber \\
&& x_2= C_2^{III} + \frac{\lvert \xi_2 \rvert}{2 {\lvert \omega \rvert}}
{\lvert Z \rvert} \
\sin ({\lvert \omega \rvert}t - \phi_2).
\label{3sol-classi}
\eea
\noindent The kinetic energy term is negative-definite and consequently,
$H^{III}$ is bounded from above. It seems that such a region has never been
encountered and explored previously within the context of Hamiltonian system
with balanced loss and gain. The system governed by the Hamiltonian
$-H^{III}= H^I$ may be interpreted as that of a particle in the background of
a Euclidean metric that is subjected to an effective external magnetic field
${\lvert \omega \rvert}$. The second order decoupled equations of motion
resulting from $H^I$ and $H^{III}$ are identical, since they differ by an
overall multiplication factor of $-1$. However, the first order equations
in these two regions are not identical. This is manifested in the fact that
the time-dependent part of $x_2$ in Eqs. (\ref{sol-classi}) and
(\ref{3sol-classi}) differ by a sign. The solutions of the decoupled
second order equations are identical in Region-I and Region-III, if solved 
with identical initial conditions. It may be noted that under the transformation
$B \rightarrow -B$, ${\cal{D}} \rightarrow {\cal{D}}^T$. Thus, the
characteristic polynomial determining the solutions of Eq. (\ref{X}) for
$V=0$ is invariant under the transformation $B \rightarrow -B$.  Consequently,
even though ${\cal{M}}$ is not positive-definite for $B <0$, the solutions are
periodic.  A duality relation exists between the two Hamiltonians in Region-I
and Region-III. In particular, the equations of motion of $H^I$ with potential
${\cal{V}}$ and that of $H^{III}$ with potential $-{\cal{V}}$ are the same.
The duality relation may be used to find solutions in Region-III from that of
Region-I and the vice verse.

A comment is in order before the end of this section. The solutions
for $C \neq 0, \gamma \rightarrow 0$ can not be obtained from Eq.
(\ref{sol-classi},\ref{2sol-classi},\ref{3sol-classi}), since the limit
is singular. However, the solutions of Eq. (\ref{X}) with ${\cal{D}}$ given
by Eq. (\ref{landau-rep}) and $V=0$ may be obtained directly with a smooth
limit $\gamma \rightarrow 0$ to the solutions of system with $C \neq 0$. In
particular,
 \bea
x_1= {\lvert A \rvert} \cos (\omega t +\phi_1), \ \
x_2= \sqrt{\frac{B-C}{B+C}} \ {{\lvert A \rvert}}\sin (\omega t +\phi_2),
\eea 
\noindent where the integration constant $A={\lvert A \rvert} e^{i \phi_1}$
and the phase $\phi_2$ has the expression:
\be
\phi_2=tan^{-1}\left ( \frac{\omega \sin \phi_1 - 2 \gamma \cos \phi_1}{
\omega \cos \phi_1 + 2 \gamma \sin \phi_1} \right ).
\ee
\noindent The center of the cyclotron motion is chosen to be at the origin
by taking the remaining two integration constants equal to zero. The motion
of the particle is confined along along $x_1$ for $B = C$ and $x_2$ diverges
is for $B=-C$. Both of these cases $B=\pm C$ belong to Region-II, in which
the solutions are diverging. The phases $\phi_1$ and $\phi_2$ become identical
in the limit of vanishing gain and loss terms, i.e. $\gamma=0$. However, the
amplitudes for the periodic solutions of $x_1$ and $x_2$ are different, leading
to elliptical orbits. The solutions of the standard Landau Hamiltonian with
circular orbits is recovered with a further choice of $C=0$. It may be noted
that the limit $C \rightarrow 0, \gamma \rightarrow 0$ is also well-defined
and independent of the order in which the limit has been taken.

\subsection{Quantum system}

The canonical quantization scheme is followed with $\hbar=1$ and the
classical variables are treated as operators satisfying the relations:
\be
\left [ x_i, x_j \right ]=0, \ \left [ p_i, p_j \right ]= 0,
\ \left [ x_i, p_j \right ]= i \delta_{ij}.
\label{com-1}
\ee
\noindent The canonical co-ordinate transformations (\ref{trans-cor}) and
Eq. (\ref{com-1}) lead to the Heisenberg algebra in the new co-ordinate
system: 
\be
\left [ {\cal{X}}_i, {\cal{X}}_j \right ]=0, \ \left [ {\cal{P}}_i,
{\cal{ P}}_j \right ]= 0,
\ \left [ {\cal{X}}_i, {\cal{P}}_j \right ]= i \delta_{ij}.
\ee
\noindent The generalized momenta $\hat{\Pi}$ and commutation relation
between its two components read, 
\be
\hat{\Pi}_1= {\cal{P}}_1 + \frac{\lvert \omega \rvert}{4} {\cal{X}}_2, \
\hat{\Pi}_2= {\cal{P}}_2 -  \frac{\lvert \omega \rvert}{4} {\cal{X}}_1, \
\left [ \hat{\Pi}_1, \hat{\Pi}_2 \right ] = i \frac{\lvert \omega \rvert}{2}.
\label{pipi}
\ee
\noindent  It is known\cite{ds-pkg1,p6-deb} for $B=0=C$ that balanced
loss-gain systems can be formulated in terms of either Landau or symmetric
gauge for the gauge potential giving rise to analogous magnetic field. The
same can be generalized\footnote{It is immaterial whether $C=0$ or not.} 
for $B\neq0$ by taking the gauge potentials leading to real and analogous
fields in the same gauge. Both the gauge potentials in Eq. (\ref{pipi}) are
taken in the symmetric gauge.

It was shown that the system defined by Eq. (\ref{H}) has an effective
description in terms of ${{H}}^a$ in Eq. (\ref{hami-eq}). These two
Hamiltonians are related by a canonical co-ordinate transformation
defined in Eq. (\ref{trans-cor}), which consists of a rotation from
$X$ to $\hat{X}$, followed by a non-uniform scale transformation,
$\hat{X}  \rightarrow {\cal{X}}$. Thus, the eigenstates $\phi(x_1,x_2)$ in
the $X$ co-ordinates are related to the eigenstates $\psi({\cal{X}}_1,
{\cal{X}}_2)$ in the ${\cal{X}}$ co-ordinates via a unitary transformation
generated by the canonical co-ordinate transformations (\ref{trans-cor}). 
In particular,
\bea
&& \phi(x_1,x_2) =  e^{i \hat{\cal{S}}} e^{ i \theta {\cal{J}}_3} \ 
\psi({\cal{X}}_1, {\cal{X}}_2),\nonumber \\
&& \hat{\cal{S}}:= \sqrt{\lambda_+} \left \{ {\cal{X}}_1,
{\cal{P}}_1 \right \}
+ \sqrt{\lambda_-} \left \{ {\cal{X}}_2, {\cal{P}}_2 \right \}, \
{\cal{J}}_3:={ {\cal{X}}_1 {\cal{P}}_2 - {\cal{X}}_2 {\cal{P}}_1 }.
\eea
\noindent The operator ${\cal{J}}_3$ is the generator of rotation around an
axis perpendicular to the `${\cal{X}}_1-{\cal{X}}_2$'-plane. The operator
${\cal{\hat{S}}}$ generates scaling for ${\cal{X}}_1$ by an amount
${\lambda_+}^{-\frac{1}{2}}$, for ${\cal{X}}_2$ by an amount 
${\lambda_-}^{-\frac{1}{2}}$. Similarly, it generates scaling for
${\cal{P}}_1$ by an amount $\sqrt{\lambda_+}$ and for ${\cal{P}}_2$ by
an amount $\sqrt{\lambda_-}$. The expectation value of an operator may
be calculated either in terms of $\psi$ and $\phi$, since they are related
through a unitary transformation. It should be noted that the unitary
transformation is allowed even for non-vanishing potential, i.e. $V \neq 0$
and without any pre-specified symmetry on it. In particular, neither
${\cal{J}}_3$ nor $\hat{\cal{S}}$ is a symmetry of the system. Thus, an exactly
solvable model in any of the two co-ordinate systems is also exactly
solvable in the other co-ordinate system with seemingly different potentials. 

\subsubsection{Region-I}

The matrix $\eta^a=I_2$ and $H$ reduces to the Landau Hamiltonian in the
symmetric gauge.
The eigen-value problem of the Landau Hamiltonian is well known\cite{landau}.
In particular, an operator $a$ and its adjoint $a^{\dagger}$ are defined as,
\be
a:= \frac{1}{ {\sqrt{\lvert \omega \rvert}}} \left ( \hat{\Pi}_1 +
i \hat{\Pi}_2, \right ), \
a^{\dagger}:= \frac{1}{ \sqrt{\lvert \omega \rvert}} \left ( \hat{\Pi}_1 -
i \hat{\Pi}_2, \right ), \
\left [ a , a^{\dagger} \right ] = 1, 
\ee
\noindent where the commutation relation between them allows to identify
these two operators as annihilation and creation operators, respectively.
The Hamiltonian and the eigenvalues can be expressed as,
\be
H^{(I)}= {\lvert \omega \rvert} \left ( a^{\dagger} a + \frac{1}{2} \right ),
\ \
E_n^{(I)}= (n+\frac{1}{2}){\lvert \omega \rvert}, \  n \in \mathbb{Z}^*.
\ee
\noindent The degenerate ground-state wave-functions for $C=0$ in terms of
the co-ordinates $(x_1, x_2)$ have the form:
\be
\phi(x_1, x_2)= \left ( \xi x_1 - \xi^* x_2 \right )^m
e^{- \frac{\lvert \omega \rvert}{8} \lvert \xi x_1 -
\xi^* x_2 \rvert^2 }, \ 
\xi=\sqrt{\frac{2}{{\lvert \omega \rvert}}}
\left ( \sqrt{{\lvert \lambda_+\rvert}}+
i \sqrt{{\lvert \lambda_-\rvert}} \right ), \ 
m \in \mathbb{Z}^*,
\label{wave-f}
\ee
\noindent where $\xi^*$ is the complex conjugate of $\xi$. Different values
of $m$ correspond to linearly independent wave-functions $\phi(x_1,x_2)$
spanning the degenerate sub-space, since $E_n^{(I)}$ is independent of $m$.
The most probable distribution of ${\lvert \phi(x_1,x_2) \rvert}^2$
is centered around an ellipse instead of a circle. The effect of the balanced
loss and gain is to distort the circle around which the most probable
distribution of the probability density occurs for the ground-state
of the Landau Hamiltonian. This is consistent with the classical result.
The annihilation and creation operators,
when expressed in terms of $(x_1, x_2, p_1, p_2)$ have the following
forms:
\be
a=\frac{1}{2} \left [ \xi^* \left ( p_1 + \frac{1}{2} x_2 \right )
+ \xi \left ( p_2 - \frac{1}{2} x_1 \right ) \right ], \ \
a^{\dagger}=\frac{1}{2} \left [ \xi \left (
p_1 + \frac{1}{2} x_2 \right ) + \xi^* \left (
p_2 - \frac{1}{2} x_1 \right ) \right ].
\ee
\noindent The excited states may be obtained by successive
operations of the operator $a^{\dagger}$ on $\phi(x_1,x_2)$.
One might define at this point a second set of creation and
annihilation operators,
\be
b=\frac{1}{2} \left [ \xi \left ( p_1 - \frac{1}{2} x_2 \right )
+ \xi^* \left ( p_2 + \frac{1}{2} x_1 \right ) \right ], \ \
b^{\dagger}=\frac{1}{2} \left [ \xi^* \left (
p_1 - \frac{1}{2} x_2 \right ) + \xi \left (
p_2 + \frac{1}{2} x_1 \right ) \right ],
\ee
\noindent which satisfy the commutation relation $ [b,b^{\dagger}]=1$.
It may be noted that $[a,b]=[a,b^{\dagger}]=0$ and similarly,
$[a^{\dagger}, b^{\dagger}]= [a^{\dagger}, b]=0$ implying that
$[H^I,b]=[H^I,b^{\dagger}]=0$. The operator ${\cal{J}}_3$,
when expressed in the original co-ordinate, has the following expression:
\bea
{\cal{J}}_3 & = & b^{\dagger} b - a^{\dagger} a\nonumber \\
& = & \frac{1}{2} {\lvert \xi \rvert}^2 J_3 + \frac{1}{4}
\left ( \xi^2 + {\xi^*}^2 \right ) \left ( x_1 p_1 -x_2 p_2 \right ),
\eea
\noindent where $J_3:=x_1 p_2 - x_2 p_1$ is the angular momentum operator
in the original co-ordinate. The Hamiltonian $H$ is not invariant under a
rotation on the $x_1-x_2$-plane due to the simultaneous presence of gauge
potentials corresponding to the `anomalous magnetic field' as well as external
magnetic field. Thus, ${\cal{J}}_3$ and $J_3$ are not identical. The operator
${\cal{J}}_3$ satisfies the following commutation relations:
\be
[H^I, {\cal{J}}_3]=0, \ \ [{\cal{J}}_3,b]=-b, \ [{\cal{J}}_3,b^{\dagger}]=
b^{\dagger}, [{\cal{J}}_3, a]= a, \ [{\cal{J}}_3,a^{\dagger}]=-
a^{\dagger}.
\ee
\noindent The operator $b^{\dagger}$ acting on the wave-function (\ref{wave-f})
increase the angular-momentum eigen-value by one unit without changing the
energy eigenvalue. On the other hand, the operator $a^{\dagger}$s increases
the energy eigen-value by one unit, while decreases the angular momentum
eigen-value by one unit. Similarly, $b$ decreases the angular momentum
eigen-value by one unit without changing the energy eigen-value, while
$a$ decreases the energy eigen-value by one unit and increases the angular
momentum eigen value by one unit. The number of degenerate states within
a finite geometry is proportional to the magnetic flux piercing through the
area.  Thus, the number of degenerate states for a fixed geometry gets reduced
in presence of the loss-gain terms.

\subsubsection{Region-II} The Hamiltonian is not positive-definite.
The analysis of the system in this region is  well suited in the Landau
gauge. The Hamiltonian in the Landau gauge may be obtained via appropriate
unitary transformation\cite{ds-pkg1}. In particular, a translational invariant
Hamiltonian $H_{L_1}$ along the $x_2$ direction may
be obtained as,
\bea
H_{L_1} &= & 
{\cal{G}}^{-1} H^{II} {\cal{G}}=
{\cal{P}}_1^2 - \left ({\cal{P}}_2 - \frac{\lvert \omega \rvert}{2}
{\cal{X}}_1 \right )^2, \ \
{\cal{G}}=e^{-i\frac{\lvert \omega \rvert}{4} {\cal{X}}_1 {\cal{X}}_2}.
\eea
\noindent Similarly, a translational invariant Hamiltonian along the
${\cal{X}}_1$ direction may also be obtained. The method of separation
of variables may be used to cast the eigen-value problem solely in terms
of ${\cal{X}}_1, {\cal{P}}_1$. In particular,
\be
\psi_0^{-1} H_{L_1} \psi_0=
{\cal{P}}_1^2 - \frac{{\lvert \omega \rvert}^2}{4} \left ({\cal{X}}_1-
\frac{2 k_2}{\lvert \omega \rvert} \right )^2, \ \
\psi_0=e^{i k_2 {\cal{X}}_2}.
\ee
\noindent The original eigenvalue-problem reduces to that of a particle
moving in an inverted oscillator potential with shifted origin. There are
no bound states.

\subsubsection{Region-III} The Hamiltonian $H^{III}=-H^{I}$ is negative
definite and bounded from above. Thus, the eigen-value problem for
$-H^{III}$ and $H^I$ are identical.

\subsection{Hall Effect}

The discussions so far have been confined to Landau Hamiltonian with a
vanishing potential $V$. A description of the Hall effect requires an
external uniform electric field. The potential $V$ is chosen as,
\bea
V(x_1,x_2)= - \frac{E}{ \omega^2} \left ( B x_1 - \gamma x_2
\right ), \ \ \omega^2= \left ( B^2 - \gamma^2 \right ). 
\eea
\noindent The coupling of the velocity dependent non-Lorentzian interaction
$C$ is taken to be zero for simplicity. The matrix ${\cal{M}}$ is singular for
$B =\pm \gamma$ and the Hamiltonian formulation is based on the assumption of
non-singular ${\cal{M}}$. Thus, the condition $B \neq \pm \gamma$ should be
imposed in order to make it consistent with the Hamiltonian formulation. The
Eq. (\ref{X}) for this choice of $V$ and $C$ has the expression,
\be
\ddot{x}_1 + \gamma \dot{x_1} - B \dot{x}_2 =E, \ \
\ddot{x}_2 - \gamma \dot{x_2} + B \dot{x}_1 =0, \ \
\label{hall-classi}
\ee
\noindent where the representation of ${\cal{M}}$ and ${\cal{D}}$ are given
by Eq. (\ref{landau-rep}). Eq. (\ref{hall-classi}) describes Hall effect
with balanced loss and gain, where $E$ is the magnitude of the external
electric field along positive $x_1$-direction. A transformation
$E \rightarrow -E$ allows to flip the direction of the electric field
to the negative $x_1$-direction. The analysis of the system is presented
in Region-I only.

The equations of motion in the transformed co-ordinate system have the form,
\bea
&& \ddot{\cal{X}}_1 - {\lvert \omega \rvert} \dot{\cal{X}}_2 ={\cal{E}}_1, \ 
\ddot{\cal{X}}_2 + {\lvert \omega \rvert} \dot{\cal{X}}_1 =
{\cal{E}}_2,\nonumber \\
&& {\cal{E}}_1 \equiv \frac{E}{2\sqrt{2 \lambda_+}}, \ {\cal{E}}_2 \equiv
- \frac{E}{2\sqrt{2 \lambda_-}},
\eea
\noindent implying that the effective electric field $\vec{\cal{E}}=
{\cal{E}}_1 \hat{\cal{X}}_1+ {\cal{E}}_2 \hat{\cal{X}}_2$ has
non-vanishing components along both the directions. The Hall current
is in the transverse direction to the external electric field $\vec{\cal{E}}$.
However, the inverse transformation from $({\cal{X}}_1, {\cal{X}}_2)$ to
$(x_1, x_2)$ involves unequal scaling of the co-ordinates followed by a
rotation, resulting in a Hall current along a direction making an angle 
$\phi_H=tan^{-1}(\frac{B}{\gamma})$ with the direction of the external
electric field $\vec{E}={\lvert E \rvert} \hat{x}_1$. In particular, the
solutions of Eq. (\ref{hall-classi}) are,
\bea
&& x_1= C_1^I + \frac{\lvert \xi_1 \rvert}{2 {\lvert \omega \rvert}}
{\lvert Z \rvert} \
\cos ({\lvert \omega \rvert}t - \phi_1)
- \frac{E \gamma}{\omega^2} t,\nonumber \\ 
&& x_2= C_2^I - \frac{\lvert \xi_1 \rvert}{2 {\lvert \omega \rvert}}
{\lvert Z \rvert} \
\sin ({\lvert \omega \rvert} t - \phi_2)
-\frac{EB}{\omega^2}t.
\label{hall-sol-classi}
\eea
\noindent It is interesting to note that the direction of the Hall current
is not perpendicular to the direction of the applied  electric field in
presence of balanced loss and gain. Further, the angle between the direction
of the Hall current and that of the external electric field depends on the
ratio of the applied external magnetic field and the loss-gain parameter.
It is known that similar behaviour for the Hall current is observed in plasma
when the Hall parameter is very high.

The quantum Hamiltonian in presence of the external electric field
is not rotationally invariant: \be
H={\hat{\Pi}}^T {\hat{\Pi}} + {\cal{V}}({\cal{X}}_1,{\cal{X}}_2), \
{\cal{V}}({\cal{X}}_1,{\cal{X}}_2)= -\frac{1}{2} \left (
{\cal{E}}_{1} {\cal{X}}_1
 + {\cal{E}}_{2} {\cal{X}}_2 \right ).
\ee
\noindent A rotation on the $({\cal{X}}_1-{\cal{X}}_2)$-plane by an angle
$\theta_1=-tan^{-1} (\frac{{\cal{E}}_2}{{\cal{E}}_1})$, followed by
unitary transformations cast the Hamiltonian $H$ as that of a one
dimensional harmonic oscillator with shifted origin:
\bea
H_1 & = & \psi_0^{-1} {\cal{G}}^{-1} e^{i \theta_1 {\cal{J}}_3} H
e^{-i \theta_1 {\cal{J}}_3} {\cal{G}} \psi_0 \nonumber \\
& = & {\cal{P}}_1^2 + \frac{{\lvert \omega \rvert}^2}{4} \left ({\cal{X}}_1-
\frac{2 k_2}{\lvert \omega \rvert} -
\frac{\lvert {\cal{E}} \rvert}{{\lvert \omega \rvert}^2} \right )^2-
\frac{1}{4 {\lvert \omega \rvert}^2}
\left ( {\cal{E}}^2 + 4 k_2 {\lvert \omega \rvert} 
{\lvert {\cal{E}} \rvert} \right ).
\eea
\noindent The rotation keeps the $\hat{\Pi}^T \hat{\Pi}$ term invariant, while
mixing the components of ${\vec{\cal{E}}}$ such that it has non-vanishing
components only along ${{\cal{X}}}_1$ in the rotated co-ordinate. The operator
${\cal{G}}$ transforms $H$ in rotational invariant gauge to $H$ in
Landau gauge with translational invariance along ${\cal{X}}_2$ direction.
The wave-function $\psi_0$ is used to decouple the kinetic energy part
involving ${\cal{P}}_2$. The energy eigenvalues are
\be
E_{n,k_2}=(n+\frac{1}{2}) {\lvert \omega \rvert} -
\frac{1}{4 {\lvert \omega \rvert}^2} \left ( {\cal{E}}^2 +
4 k_2 {\lvert \omega \rvert} {\lvert {\cal{E}} \rvert} \right ),
\ee
\noindent while the eigenfunctions are that of one dimensional harmonic
oscillator in terms of the shifted co-ordinate $\tilde{\cal{X}}_1={\cal{X}}_1-
-\frac{2 k_2}{\lvert \omega \rvert} - \frac{\lvert {\cal{E}} \rvert}{{\lvert
\omega \rvert}^2}$. The probability current  $\vec{{J}}$ has non-vanishing
components only along ${\cal{X}}_2$ direction, i.e. the transverse direction
to the direction of the applied electric field. In particular,
\be
{{J}}_1=0, \ \ {{J}}_2= 2 \rho \left ( k_2 - \frac{\mid \omega \mid}{2}
{\cal X}_1 \right ),
\ee
\noindent where $\rho$ is the probability density. However, when expressed
in terms of the original co-ordinates $(x_1, x_2)$, the probability current has
non-vanishing components along the directions of both $x_1$ and $x_2$,
i. e. the direction of the external electric field and its transverse
direction, respectively. This is consistent with the classical result.

\subsection{Spin, Pauli equation \& Supersymmetry}

Landau Hamiltonian with spin degrees of freedom for the particle contains
an additional term due to the interaction between its magnetic moment
and the external magnetic field. This gives rise to Pauli equation and
appears in diverse branches of physics. A remarkable property of the
Pauli equation is that it has an underlying supersymmetry\cite{khare1}.
In this section, Landau Hamiltonian with balanced loss/gain term is
generalized by including the spin degrees of freedom for the particle and
shown to admit underlying supersymmetry. The supersymmetric Hamiltonian
$H_S$ in Region-I(${\lvert B \rvert} > \gamma$) is taken as,
\bea
H_S= \frac{B}{2} \left ( \Pi_1^2 + \Pi_2^2 \right ) + \frac{\gamma}{2}
\{ \Pi_1, \Pi_2 \} + \frac{{\mid \omega \mid}}{2} \sigma_z,
\eea
\noindent where the co-efficient of the spin-dependent term is chosen
such that $H_S$ is supersymmetric. It may be noted that the Zeeman energy
contains the effective magnetic field ${\lvert \omega \lvert}=B \sqrt{1-
(\frac{\gamma}{B})^2}$ instead of the external magnetic field $B$. This
is essential in order to have underlying supersymmetry in $H_S$. This may
also be interpreted as that the spin interacts with the external magnetic
field $B$, but, the Land$\acute{e}$ $g$-factor is modified by a multiplicative
factor of $\sqrt{1- (\frac{\gamma}{B})^2}$ which reduces to its standard value
in the limit $\gamma \rightarrow 0$.
It is known that $H^{III}$ is negative-definite and does not admit
supersymmetric generalizations. Similarly, $H^{II}$ is not semi-positive
definite and excludes the possibility of supersymmetric generalizations.

The supersymmetric Hamiltonian $H_S$ can be expressed in
terms of the canonically transformed momenta $(\hat{\Pi}_1, \hat{\Pi}_2)$
as,
\bea
H_S=\hat{\Pi}_1^2 + \hat{\Pi}_2^2 + \frac{\mid \omega \mid}{2} \sigma_z,
\label{pauli}
\eea
\noindent which has the standard form of the Pauli Hamiltonian.
It is known that the Pauli Hamiltonian in Eq. (\ref{pauli}) has a 
supersymmetric factorization, which can be used to introduce
supercharges for $H_S$ in terms of $\Pi_1$ and $\Pi_2$. In particular,
\bea
&& q_1= \sigma_x \hat{\Pi}_1 - \sigma_y \hat{\Pi}_2 =
\sqrt{\frac{\lambda_+}{2}} (\Pi_1+\Pi_2) \sigma_x -
\sqrt{\frac{\lambda_-}{2}} (\Pi_1-\Pi_2) \sigma_y,\nonumber \\
&& q_2= \sigma_x \hat{\Pi}_2 + \sigma_y \hat{\Pi}_1 =
-\sqrt{\frac{\lambda_-}{2}} (\Pi_1-\Pi_2) \sigma_x +
\sqrt{\frac{\lambda_+}{2}} (\Pi_1+\Pi_2) \sigma_y
\eea
\noindent which satisfy the relations $\{q_a, q_b \}=2 \delta_{ab} H_S, \
a, b=1, 2$. It should be mentioned here that another set of two supercharges
satisfying the relations $\{Q_a,Q_b\}=2 \delta_{ab} H_S$ may be defined as,
\bea
Q_1 & = & \left ( \sigma_x \Pi_1 + \sigma_y \Pi_2 \right ) \sqrt{\frac{B}{2}}
\cos \tilde{\theta} + \left ( \sigma_x \Pi_2 + \sigma_y \Pi_x \right )
\sqrt{\frac{B}{2}} \sin \tilde{\theta}\nonumber\\
& = & \frac{\hat{\Pi}_1}{2} \sqrt{\frac{B}{\lambda_+}} \left ( \sigma_x +
\sigma_y \right ) \left ( \cos \tilde{\theta} + \sin \tilde{\theta} \right )
-\frac{\hat{\Pi}_2}{2} \sqrt{\frac{B}{\lambda_-}} \left ( \sigma_x - 
\sigma_y \right ) \left ( \cos \tilde{\theta} - \sin \tilde{\theta} \right )
\nonumber \\
Q_2 & = & \left ( \sigma_x \Pi_1 - \sigma_y \Pi_2 \right ) \sqrt{\frac{B}{2}}
\sin \tilde{\theta} + \left ( \sigma_x \Pi_2 - \sigma_y \Pi_x \right )
\sqrt{\frac{B}{2}} \cos \tilde{\theta},\nonumber \\
& = & \frac{\hat{\Pi}_1}{2} \sqrt{\frac{B}{\lambda_+}} \left ( \sigma_x-
\sigma_y \right ) \left ( \cos \tilde{\theta} + \sin \tilde{\theta} \right )
+ \frac{\hat{\Pi}_2}{2} \sqrt{\frac{B}{\lambda_-}} \left ( \sigma_x + 
\sigma_y \right ) \left ( \cos \tilde{\theta} - \sin \tilde{\theta} \right ).
\eea
\noindent where $\tilde{\theta}=-\frac{1}{2} \tan^{-1}
\frac{\gamma}{{\mid \omega \mid}}$. It should be noted here that all the four
supercharges $(Q_1, Q_2, q_1, q_2)$ taken together do not give rise to
${\cal{N}}=4$ supersymmetry. Each set of supercharges $(Q_1, Q_2)$ and
$(q_1,q_2)$ corresponds to ${\cal{N}}=2$ supersymmetry only. The factorization
of the Hamiltonian is achieved in two different ways. The eigen-value problem
and the state-space structure of $H_S$ in Eq. (\ref{pauli}) is well
known\cite{khare1} and is not reproduced here.

 \section{Summary \& Discussions} 
It has been shown that the kinetic energy term for the Hamiltonian systems
with balanced loss and gain may be made to be positive-definite by including
a Lorentz interaction in the system. The result is quite general and applicable
to a large class of systems with space-dependent loss-gain terms. A few
representations of the matrices appearing in the definition of the Hamiltonian
is given with the identification of the regions in the parameter space of
the theory in which the kinetic energy term is positive-definite. The presence
of Lorentz interaction allows the balancing of loss-gain terms to occur in as
many ways as the solutions of $Tr({\cal{D}})=0$ can be realized. The pair-wise
balancing of loss-gain terms, which is a necessity in absence of the Lorentz
interaction, appears as a special case in the present situation.

One important aspect of the present formulation is that the Hamiltonian
system with balanced loss and gain may be interpreted as defined in the
background of a metric without any loss-gain terms. The absence of loss-gain
terms is manifested in modifying the magnitude of the external magnetic field
due to the Lorentz force as well as coupling constants of various
velocity-dependent non-Lorentzian interaction terms. The specific signature
of the background metric depends on the form of the Hamiltonian defining
the system. The effective Hamiltonian for the case of background Euclidean
metric contains only Lorentz interaction with modified magnitude.
 
The classical and the quantum Landau Hamiltonian with balanced loss and gain
have been studied in some detail. There are three regions in the parameter
space depending on the nature of the background metric in the effective
description of the system. The background metric is Euclidean provided the
magnitude of the effective magnetic field is less than the magnitude of the
applied magnetic field and the metric is pseudo-Euclidean, otherwise. 
The classical equations of motion are solved exactly in all three regions
with periodic solutions in Region-I and Region-III, which correspond to
positive-definite and negative-definite background metric, respectively. In
these two regions, the particle moves in an elliptic
orbit with a cyclotron frequency that is less than its value in absence
of loss-gain terms. There are no periodic solutions in Region-II corresponding
to a background pseudo-Euclidean metric. The quantum bound states are obtained
in Region-I and Region-III, consistent with the classical description.
The results of the standard Landau Hamiltonian are valid even in presence of
balanced loss and gain with a reduced value of the cyclotron frequency in
these two regions. The Region-II does not admit any bound state.

The Hall effect with balanced loss and gain has been studied by including
an external uniform electric field to the Landau Hamiltonian. One very
interesting result for this case is that the Hall current is not necessarily
in the perpendicular direction to the applied external electric field. The
Hall current has non-vanishing components along the direction of the
external electric field as well as to its transverse direction. This result
is valid at the classical as well as quantum level. 
Similar results are known
to exist in plasma in case the Hall parameter, the ratio between electron
cyclotron frequency and the electron-heavy-particle collision frequency, is
high. For the case of balanced loss-gain system, the Hall angle depends
on the ratio of the external magnetic field and the gain-loss parameter.
Any possible connection between these two systems is worth exploring in
future.

The Pauli equation with balanced loss and gain has been studied from
the viewpoint of underlying supersymmetry in the system. It has been
shown that the system admits ${\cal{N}}=2$ supersymmetry, if the Zeeman
energy contains the effective magnetic field ${\lvert \omega \rvert}$ instead
of the external magnetic field $B$. This Zeeman energy term may also
be interpreted as interaction between the external magnetic field $B$
and the spin degrees of freedom, but, with a modified Land$\acute{e}$
$g$-factor due to the presence of
loss-gain terms. The state-space structure and the spectra are identical
with the standard Pauli Hamiltonian. However, inclusion of Dresselhaus
and Rashba spin-orbit interactions in the Hamiltonian is expected to give
significant results and such investigations will be pursued in future.

\section{Acknowledgments}

This work is partly supported by a grant ({\bf SERB Ref. No. MTR/2018/001036})
from the Science \& Engineering Research Board(SERB), Department of Science
\& Technology, Govt. of India under the {\bf MATRICS} scheme.

\section{Appendix-A: Representation of ${\cal{M}}$, $R$ and
${\cal{D}}$}

Several representations of ${\cal{M}}$, $R$ and ${\cal{D}}$ for the case of
pair-wise balancing of loss and gain terms have been discussed
in the main text. The purpose of this Appendix is to present two
different representations of  ${\cal{M}}$, $R$ and ${\cal{D}}$, where
the balancing of loss-gain terms are not necessarily in a pair-wise
fashion.
\subsection{Representation-I}

The $N \times N$ symmetric matrix ${\cal{M}}$ is given by,
\bea
{\cal{M}}= p\ I_N + q \ T, \ \ [T]_{ij}= \delta_{i+1,j} + \delta_{i,j+1},
\ p, q \in \Re.
\eea
\noindent The matrix ${\cal{M}}$ has the eigenvalues,
\be
\lambda_k=p + 2 q \cos (\frac{k \pi}{N+1}), k=1, 2, \dots, N, 
\ee
\noindent and is positive-definite for $p > 2 {\lvert q \rvert}$.
The matrix $\hat{O}$ that diagonalizes ${\cal{M}}$ has the expression,
\be
[\hat{O}]_{ij} =\sqrt{\frac{2}{N+1}} \sin(\frac{ij \pi}{N+1}).
\ee
\noindent For a generic choice of the matrix $J$ and the anti-symmetric matrix
$A$, $R$ is an antisymmetric matrix, $R^T=-R$. The matrix ${\cal{D}}$ has the
following expression:
\be
[{\cal{D}}]_{ij}= p \ [R]_{ij} + q \left ( [R]_{i+1,j} + [R]_{i-1,j} \right ),
\ee
\noindent where $[R]_{N+1,j}$ and $[R]_{0,j}$ are taken to be zero. 
It is apparent from the expressions of the diagonal elements
$[{\cal{D}}]_{i,i}= q ( [R]_{i+1,i} + [R]_{i-1,i})$ and $[R]_{ij}=-[R]_{ji}$
that the balancing of loss/gain terms does not necessarily occur in a
pair-wise fashion. This particular representation with $V=\frac{1}{2} X^T X$
corresponds to a chain of linear oscillators with nearest-neighbour
interaction and balanced loss-gain that is subjected to velocity mediated
coupling among different degrees of freedom.

\subsection{Representations-II} 

The $N \times N$ symmetric matrix ${\cal{M}}$ with its elements
$[{\cal{M}}]_{ij}$ given by,
\bea
[{\cal{M}}]_{ij}= p \delta_{ij} + q \left ( 1 - \delta_{i,j} \right ),
\label{matm}
\eea
\noindent has eigenvalues $p-q$ with multiplicity $N-1$ and $p+(N-1) q$. 
The matrix is positive-definite provided,
\be
p > 0, \ \ -\frac{p}{N-1} < q < p.
\ee
\noindent For a generic antisymmetric matrix $R$, the matrix ${\cal{D}}$ has
the expression:
\bea
[{\cal{D}}_S]_{ij} = \frac{q}{2} \sum_{k=1}^N \left (
[R]_{kj} + [R]_{ki} \right ), \ \
[{\cal{D}}_A]_{ij}= (p+q) [R]_{ij} + \frac{q}{2} \sum_{k=1}^N \left ( [R]_{kj}
- [R]_{ki} \right ). 
\eea
\noindent The balancing of gain-loss terms does not necessarily occur in a
pair-wise fashion for this particular representation.

\end{document}